\documentclass[twoside,twocolumn,english,prl, showpacs]{revtex4}
\usepackage[T1]{fontenc}
\usepackage[latin9]{inputenc}
\usepackage{amsbsy}
\usepackage{graphicx}
\usepackage{amssymb}

\makeatletter
\@ifundefined{textcolor}{}
{%
 \definecolor{BLACK}{gray}{0}
 \definecolor{WHITE}{gray}{1}
 \definecolor{RED}{rgb}{1,0,0}
 \definecolor{GREEN}{rgb}{0,1,0}
 \definecolor{BLUE}{rgb}{0,0,1}
 \definecolor{CYAN}{cmyk}{1,0,0,0}
 \definecolor{MAGENTA}{cmyk}{0,1,0,0}
 \definecolor{YELLOW}{cmyk}{0,0,1,0}
 }

\makeatother

\usepackage{babel}

\begin{document}

\title{Sign reversal of the boson-boson interaction potential for the planar
Bose-Fermi mixtures under synthetic magnetic field }

\author{T. P. Polak}

\address{Faculty of Physics, Adam Mickiewicz University of Pozna\'{n}, Umultowska
85, 61-614 Pozna\'{n}, Poland}
\begin{abstract}
We study the mutually coupled, strongly interacting bosonic and non-interacting
fermionic, species of unequal masses in the regime were the retardation
effects are an important part of the physics. A cloud of neutral atoms
experiences a synthetic magnetic field because of a vector potential
that imposes a phase shift on the constituents. The magnetic field
causes the oscillations of the magnitude and sign of the effective
interaction between bosons from repulsive to attractive in contrast
to the static case. We show that the dynamics for the gaseous Bose-Fermi
mixtures when reaching the quantum-Hall regime becomes highly nontrivial.
\end{abstract}

\pacs{05.30.Jp, 03.75.Lm, 03.75.Nt}

\maketitle
The ultracold gases of atoms allow for the observation and control
of many-body quantum effects at macroscopic scales. They can thus
offer a fertile arena for exploration of condensed matter physics.
The obvious limitations that come from neutrality of atoms preclude
the observation of great variety of fundamental phenomena i.e. charge
particle moving in magnetic field. However, the analysis of the latter
can be carried considering rotating Bose-Einstein condensates \cite{coddington,tung,schweikhard}
trapped in lattice potential created by lasers. In a frame of reference
rotating about the $z$-axis with angular velocity $\Omega$ the kinetic
term in Hamiltonian is equivalent to that of a particle of charge
$Q$ experiencing a magnetic field $B$ with $QB=2m\Omega$, where
$m$ is the mass of the particle \cite{bhat2}. This connection shows
that the Coriolis force in the rotating frame plays the same role
as the Lorentz force on a charged particle in an uniform magnetic
field \cite{cooper,leggett}. The above setting comes with limitations
because a large magnetic fields $f\equiv ma^{2}\Omega/\pi\hbar$ (angular
momentum) are required to make possible the study of poorly explored
bosonic states in case when $f\equiv p/q$, ($p$ and $q$ is the
ratio of atom number to the number of flux quanta respectively) is
a rational number. Very recently Lin \emph{et al.} \cite{lin} circumvented
the problem by imprinting a quantum mechanical phase on the neutral
atoms without rotating them at all and the quantum-Hall physics can
be reached.

Current experiments \cite{catani,ospelkaus,ferlaino,gunter,best}
on trapped mixtures of the atomic Bose-Fermi (BF) and Bose-Bose gases
show that the presence of a relevant fraction of one modifies the
quantum phase transition occurring in\emph{ }the other inducing a
significant loss of coherence. These observations are supported by
a theoretical description that includes the multiband virtual transitions
\cite{lutchyn}, different masses of strongly interacting particles
\cite{polak2} and numerical calculations \cite{buonsante}. The density-density
(DD) interaction between different species can be repulsive or attractive
and is produced by changes of one species density that induce a modulation
of another. Therefore the dynamics underlying the phase transitions
in the BF mixtures is produced by the feedback of the density perturbation
and a shift of the inter-bosonic potential occurs, that changes the
original interaction between them \cite{mazzarella} providing various
novel phases \cite{illuminati}.

In the present paper, motivated by recent experiments done by Lin
\emph{et al.} \cite{lin}, we calculated the form of the effective
inter-bosonic potential when a synthetic magnetic field (SMF) is applied
to neutral gaseous Bose-Fermi mixtures. We predict that the fermion-mediated
effective interaction between bosons has a complicated pattern of
the frequency dependent magnitude. Moreover, the SMF renders the inter-bosonic
potential oscillatory with sign change, thus switching it between
repulsive and attractive. As a consequence the resonances appear and
BF mixture that enters the quantum-Hall regime displays surprisingly
complex dynamics unreachable in conventional solid state physics.
We expect that our theoretical results open up the experimental studies
\cite{will} of the renormalized interaction energies in stable many-body
phases with strong correlations and their dynamical properties.

Restricting our analysis to the lowest energy band of a square optical
lattice in synthetic magnetic field, the Bose-Fermi quantum gaseous
mixture can be modeled via the following Hamiltonian \cite{albus}:
\begin{eqnarray}
\mathcal{H} & = & \frac{U_{b}}{2}\sum_{i}n_{bi}\left(n_{bi}-1\right)-\sum_{\left\langle i,j\right\rangle }t_{bij}b_{i}^{\dagger}b_{j}-\mu_{b}\sum_{i}n_{bi}\nonumber \\
 & - & \sum_{\left\langle i,j\right\rangle }t_{fij}c_{i}^{\dagger}c_{j}-\mu_{f}\sum_{i}n_{fi}+U_{bf}\sum_{i}n_{bi}n_{fi},\label{hamiltonian}\end{eqnarray}
where $b_{i}^{\dagger}$($c_{i}^{\dagger}$) and $b_{j}$($c_{j}$)
stand for the bosonic (fermionic) creation and annihilation operators;
$n_{bi}=b_{i}^{\dagger}b_{i}$($n_{fi}=c_{i}^{\dagger}c_{i}$) measures
the corresponding boson (fermion) number on the site $i$, $U_{b}>0$
is the on-site repulsion and $\mu_{b}$$\left(\mu_{f}\right)$ stands
for the chemical potential for bosons (fermions). The DD interaction
between the bosonic and non-interacting, spin-polarized (collisions
in the $s$-wave channel are forbidden by their statistics), fermionic
atoms is denoted by $U_{bf}$ and depends, on boson to fermion mass
ratio $m_{b}/m_{f}$. Here, $\left\langle i,j\right\rangle $ identifies
summation over the nearest-neighbor sites. Furthermore $t_{b}$($t_{f}$)
sets the kinetic energy scale for bosons (fermions). 

A synthetic magnetic field $\boldsymbol{B}=\nabla\times\boldsymbol{A}\left(\boldsymbol{r}\right)$
enters the Hamiltonian Eq. (\ref{hamiltonian}) through the Peierls
phase factor according to $t_{ij}\rightarrow t_{ij}\exp\left(\frac{2\pi i}{\Phi_{0}}\int_{\boldsymbol{r}_{j}}^{\boldsymbol{r}_{i}}\boldsymbol{\mathrm{A}}\cdot d\boldsymbol{l}\right)$,
where $\Phi_{0}=hc/e$ is the flux quantum and $e$ is an elementary
charge. Thus, the phase shift on each site is determined by the vector
potential $\boldsymbol{\mathrm{A}}\left(\boldsymbol{r}\right)$ and
can be controlled experimentally \cite{lin}. The magnetic field is
introduced in the theory by the density of states (DOS). There are
significant difficulties in obtaining and analyzing and analyze the
solutions of the above analytically for every value of $f$. Only
a few closed formulas for DOS are accessible \cite{polak1} and consequently
not every applied magnetic field can be described theoretically. We
expanded the set of the available analytical solutions obtaining closed
formulas also for $f=1/8$ and $3/8$. This allows us the detailed
analysis of the dynamical response function which have been found
to play a crucial role in complex systems. The partition function
of bosonic and fermionic mixtures is written in the form $\mathcal{Z}=\int\left[\mathcal{D}\bar{b}\mathcal{D}b\mathcal{D}\bar{c}\mathcal{D}c\right]e^{-\mathcal{S}\left[b,c\right]}$
with action given by\begin{equation}
\mathcal{S}=\mathcal{S}_{b}+\mathcal{S}_{c}+\int_{0}^{\beta}d\tau\mathcal{H\left(\tau\right)}\end{equation}
where $\mathcal{S}_{b}=\sum_{i}\int_{0}^{\beta}d\tau\bar{b}_{i}\frac{\partial}{\partial\tau}b_{i}$
and $\mathcal{S}_{c}=\sum_{i}\int_{0}^{\beta}d\tau\bar{c}_{i}\frac{\partial}{\partial\tau}c_{i}$.
Using the bosonic (fermionic) path integral over the complex fields
depending on the {}``imaginary time'' $0\leq\tau\leq\beta\equiv1/k_{\mathrm{B}}T$
with $T$ being the temperature we can easily integrate over the fermionic
fields \cite{polak2} because spins are frozen due to influence of
the magnetic trap and there is no direct interaction between fermions.
After that, we obtain the partition function in the form $\mathcal{Z}=\int\left[\mathcal{D}\bar{b}\mathcal{D}b\mathcal{D}\bar{c}\mathcal{D}c\right]e^{-\mathcal{S}_{b}\left[b,n_{b}\right]}e^{-\mathrm{Tr}\ln\hat{G}_{c}}$.
The trace of the two-point correlation function for noninteracting
fermions $\hat{G}_{c}$, after exploiting Fourier-Matsubara transform
reads:\begin{equation}
\mathrm{Tr}\ln\hat{G}_{c}=-\frac{U_{bf}^{2}}{2}\sum_{\boldsymbol{k},\ell}\Lambda_{\boldsymbol{k}}\left(\omega_{\ell}\right)\chi_{\boldsymbol{k}}\left(i\nu_{\ell}\right)\Lambda_{-\boldsymbol{k}}\left(-\omega_{\ell}\right),\label{trace}\end{equation}
where $\omega_{\ell}=2\pi\ell/\beta$ ($\nu_{\ell}=\pi\left(2\ell+1\right)/\beta$)
with $\ell=0,\pm1,\pm2,...$ are the Bose(Fermi)-Matsubara frequencies
respecting periodic (antiperiodic) boundary conditions of the bosonic
(fermionic) field operator with $\Lambda_{\boldsymbol{k}}\left(\omega_{\ell}\right)=\bar{b}_{\boldsymbol{k}}\left(\omega_{\ell}\right)b_{\boldsymbol{k}}\left(\omega_{\ell}\right)$
and \begin{equation}
\chi_{\boldsymbol{k}}\left(i\nu_{\ell}\right)=\sum_{\mathbf{k}'}\frac{n_{\mathrm{F}}\left(t_{f\boldsymbol{k}'}^{p/q}\right)-n_{\mathrm{F}}\left(t_{f\boldsymbol{k}'+\boldsymbol{k}}^{p/q}\right)}{t_{f\boldsymbol{k}'}^{p/q}-t_{f\boldsymbol{k}'+\boldsymbol{k}}^{p/q}-i\nu_{\ell}},\end{equation}
is the Lindhard function - more commonly called the random phase approximation
with $n_{\mathrm{F}}(x)$ being the Fermi distribution; $t_{\boldsymbol{k}'}^{p/q}$
is the dispersion relation calculated from Harper equation \cite{polak1}.
It correctly predicts a number of properties of the collective phenomena
in electron gas such as plasmons \cite{giuliani}. To stay in the
local regime we perform $\boldsymbol{k}$ and \textbf{$\boldsymbol{k}'$}
integration over the first Brillouin zone and, in the $T\rightarrow0$
limit, using an analytic continuation, we obtain imaginary part $\chi''\left(\omega\right)$
of the local dynamic Lindhard function (LDLF). Therefore, the corresponding
real part $\chi'\left(\omega\right)$ can be deduced from the Kramers-Kr\"{o}nig
relation. We noticed that $\chi''\left(\omega\right)$ is proportional
to the absorption spectrum of the medium so it can be directly measured.
Now, doing the inverse Fourier transform Eq. \ref{trace} and using
gradient expansion, we obtain quadratic form of the trace with extracted
frequency dependence

\begin{equation}
\mathrm{Tr}\ln\hat{G}_{c}\rightarrow-\frac{U_{bf}^{2}\chi'\left(\omega\right)}{2}\sum_{i}\int_{0}^{\beta}d\tau\left[\bar{b}_{i}\left(\tau\right)b_{i}\left(\tau\right)\right]^{2}.\label{trace-1}\end{equation}
The consequence of the difference in masses of bosons and fermions
is the fact that the speed of the Bogoliubov sound $v_{b}$ for bosons
differs from the first sound $v_{f}$ of the ideal Fermi gas. In typical
experimental realizations the acoustic long-wavelength boson and fermion
velocities are comparable and both constituents equilibrates similarly.
The mentioned different mass ratio has far-reaching consequences,
including the possibility of generating the DD oscillations \cite{mering}.
With these concerns in mind we do not restrict our calculations to
the static limit but consider also the local dynamical response function
hence the retardation effects are an important part of the physics.
Moreover, we see \cite{polak2} from the $U_{bf}\left(m_{b}/m_{f}\right)$
dependence that even if the fermion mass is very large system has
a finite value of the interaction strength. On the other hand slow
bosonic atoms would affect the system providing strong inter-species
coupling considerably faster. We neglected the back action of the
bosons on the fermions (renormalization of the fermion DD correlation
function) because of the diluteness of the system and theory limitations
condition $U_{b}>U_{bf}^{2}\chi'\left(\omega\right)$ \cite{efremov}.

Our analysis is carried out in the frequency domain, although measurements
are sometimes made in the time domain and then Fourier transformed
to the frequency. When we add Eq. (\ref{trace-1}) to the bosonic
part of the action there is a striking resemblance to the one-component
Bose-Hubbard action with the original repulsive interaction replaced
now by%
\begin{figure}
\includegraphics[scale=0.33]{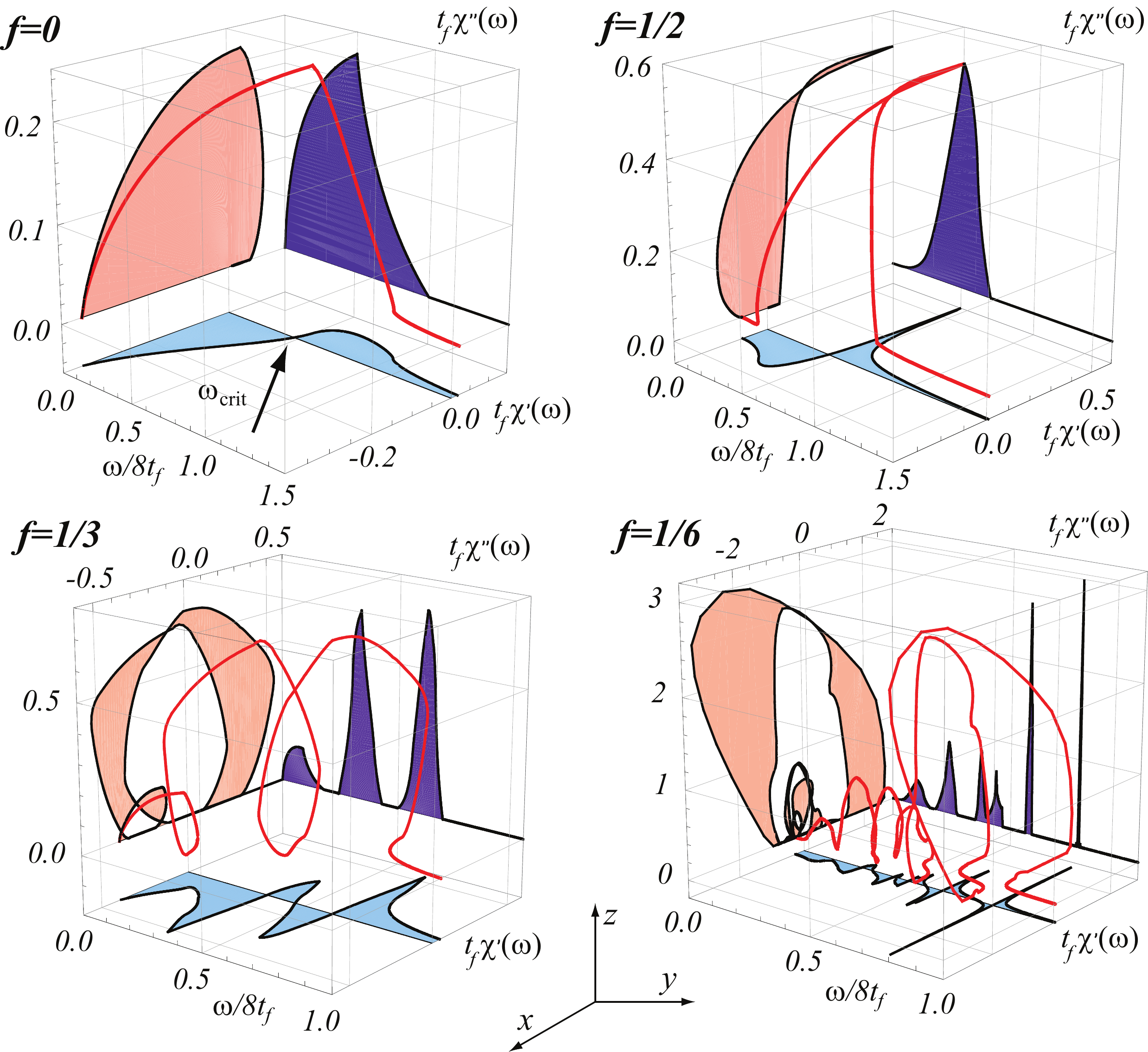}\caption{(Color online) Real $t_{f}\chi''\left(\omega\right)$ (x-y plane)
and imaginary $t_{f}\chi'\left(\omega\right)$ (z-y plane) parts of
the local, frequency $\omega/8t_{f}$ dependent, density-density response
function in the synthetic magnetic field for several values of $f=0$,
$1/2$, $1/3$ and $1/6$ (for the case with $f=1/4$ see supplemental
material \cite{supp}). The complex plot (x-z plane) is the {}``Cole-Cole''
like diagram \cite{cole}. The three-dimensional parametric curve
shows the evolution of the density-density response function with
the frequency and guarantees the stability of the system according
to the Nyquist theorem (see text). Normalization here $t_{f}\equiv t_{f}^{p/q=0}$. }
\label{Fig1}
\end{figure}
 $U_{b}\rightarrow U_{\mathrm{eff}}=U_{b}+U_{bf}^{2}\chi'\left(\omega\right)$
which is the induced, frequency-dependent, effective inter-bosonic
potential. From the above we see the DD correlations between the constituents
give rise to additional interaction among bosons, which is \emph{robust}
to repulsive or attractive nature of the inter-species interaction
but not to the sign of the LDLF. Before carrying out an analysis of
the phase diagrams it is a good idea to examine the structure of the
response function graphically Fig. \ref{Fig1}. Structure that will
usually reflect, at least in part, the physical processes present
in the real systems. The interactions caused by the DD correlations
may change its sign even if the system has not the phase shift imprinted
on it as a result of the collective excitations. For values $f=0$
and $1/2$ we have only one resonance and induced attractive interaction
between bosons becomes repulsive above some critical frequency $\omega_{\mathrm{crit}}$.
Decreasing the strength of SMF makes the physics nontrivial and for
$q>2$ the number of the resonances increase significantly Fig. \ref{Fig2}.
Such complex behavior of the mixtures emerges in the limit \emph{not
reachable} in conventional systems of condensed matter physics because
the very high values of magnetic field are required to acquire the
desired range $f\leq1/2$. The method of an optically synthesized
magnetic field for ultracold neutral atoms \cite{lin} already reached
that regime with a high accuracy. We see that the situation is greatly
modified if one applies lower magnetic fields. The narrower peaks
in frequency domain the longer time we need to observe the collective
responses from the system. According to the above we notice that the
BF mixtures has to be in a very high magnetic field to make oscillations
experimentally observable and seen as a density modulation. The stability
of the effects we analyze comes from the Nyquist theorem for the frequency
dependent diagram Fig. \ref{Fig1}. The parametric curve $\chi'\left(\omega\right)$
vs $\chi''\left(\omega\right)$ as a function of frequency and the
ensuing effective interaction described by its behavior $U_{\mathrm{eff}}$
automatically yields a stability if the curve does not encircle the
origin \cite{chavanis}.%
\begin{figure}

\includegraphics[scale=0.65]{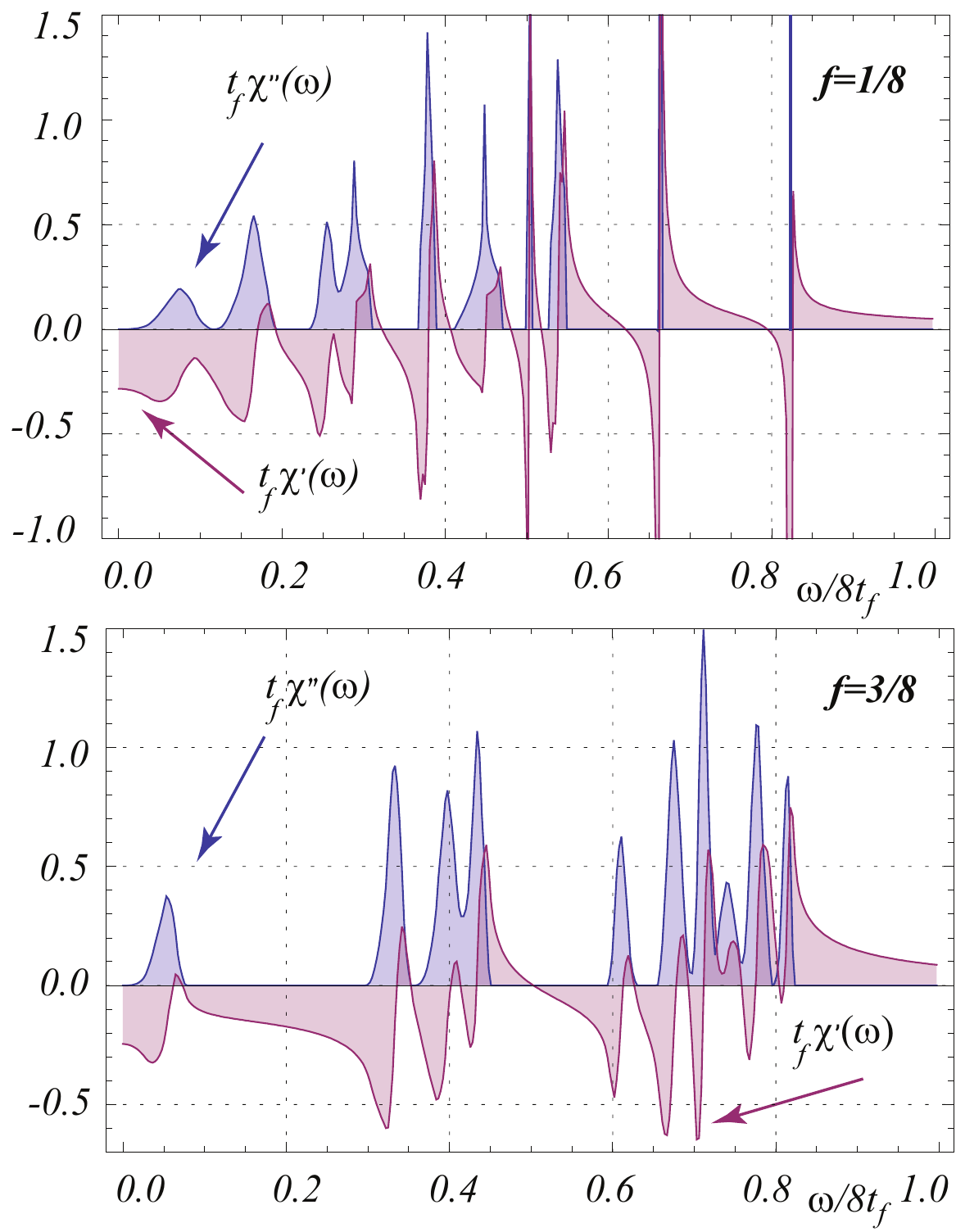}\caption{(Color online) Real $t_{f}\chi''\left(\omega\right)$ and imaginary
$t_{f}\chi'\left(\omega\right)$ parts of the local, frequency $\omega/8t_{f}$
dependent, density-density response function in the synthetic magnetic
field for $f=1/8$ and $f=3/8$. Normalization here $t_{f}\equiv t_{f}^{p/q=0}$. }
\label{Fig2}

\end{figure}

To obtain an equation of state we apply the quantum-rotor approach,
that successfully casts an essential part of the physics of strongly
interacting fermions \cite{florens,kopec,zhao} and bosons \cite{polak}
exclusively, to BF mixtures. We incorporate fully our calculations
to the phase fluctuations $b_{i}\left(\tau\right)=\left[a_{0}+a'_{i}\left(\tau\right)\right]e^{i\phi_{i}\left(\tau\right)}$
governed by the gauge $\mathrm{U}\left(1\right)$ group and drop corrections
to the amplitude $a'_{i}\left(\tau\right)$ of the order parameter
$\Psi_{\mathrm{B}}\equiv\left\langle b_{i}\left(\tau\right)e^{i\phi_{i}\left(\tau\right)}\right\rangle =b_{0}\psi_{\mathrm{B}}$.
The non-vanishing value of the $\Psi_{\mathrm{B}}$ signals a bosonic
condensation \cite{polak}. A phase $\phi_{i}\left(\tau\right)$ of
the many-body wave function might be arbitrary but correlations among
the local phases of its constituents can bring unusual gauge structures.
Now, the partition function $\mathcal{Z}=\int\left[\mathcal{D}\phi\right]e^{-\mathcal{S}_{\mathrm{ph}}\left[\phi\right]},$
with an effective action can be expressed in \emph{phase-only} terms\begin{eqnarray}
\mathcal{S}_{\mathrm{ph}}\left[\phi\right] & = & \int_{0}^{\beta}d\tau\left\{ \sum_{i}\left[\frac{\dot{\phi_{i}^{2}}\left(\tau\right)}{2U_{\mathrm{eff}}}+\frac{\bar{\mu}_{b}}{iU_{\mathrm{eff}}}\dot{\phi_{i}}\left(\tau\right)\right]\right.\nonumber \\
 & - & \left.\sum_{\left\langle i,j\right\rangle }J^{p/q}e^{i\left[\phi_{i}\left(\tau\right)-\phi_{j}\left(\tau\right)\right]}\right\} .\label{action only phase}\end{eqnarray}
The phase stiffness coefficient given by $J{}^{p/q}=t_{b}^{p/q}\left(8t_{b}^{p/q}+\bar{\mu}_{b}-U_{bf}N_{\mathrm{F}}\right)/U_{b}$
describes the hopping matrix elements renormalized by the amplitude
of the order parameter and $N_{\mathrm{F}}$ is the average number
of fermions. 

\begin{figure}
\includegraphics[scale=0.33]{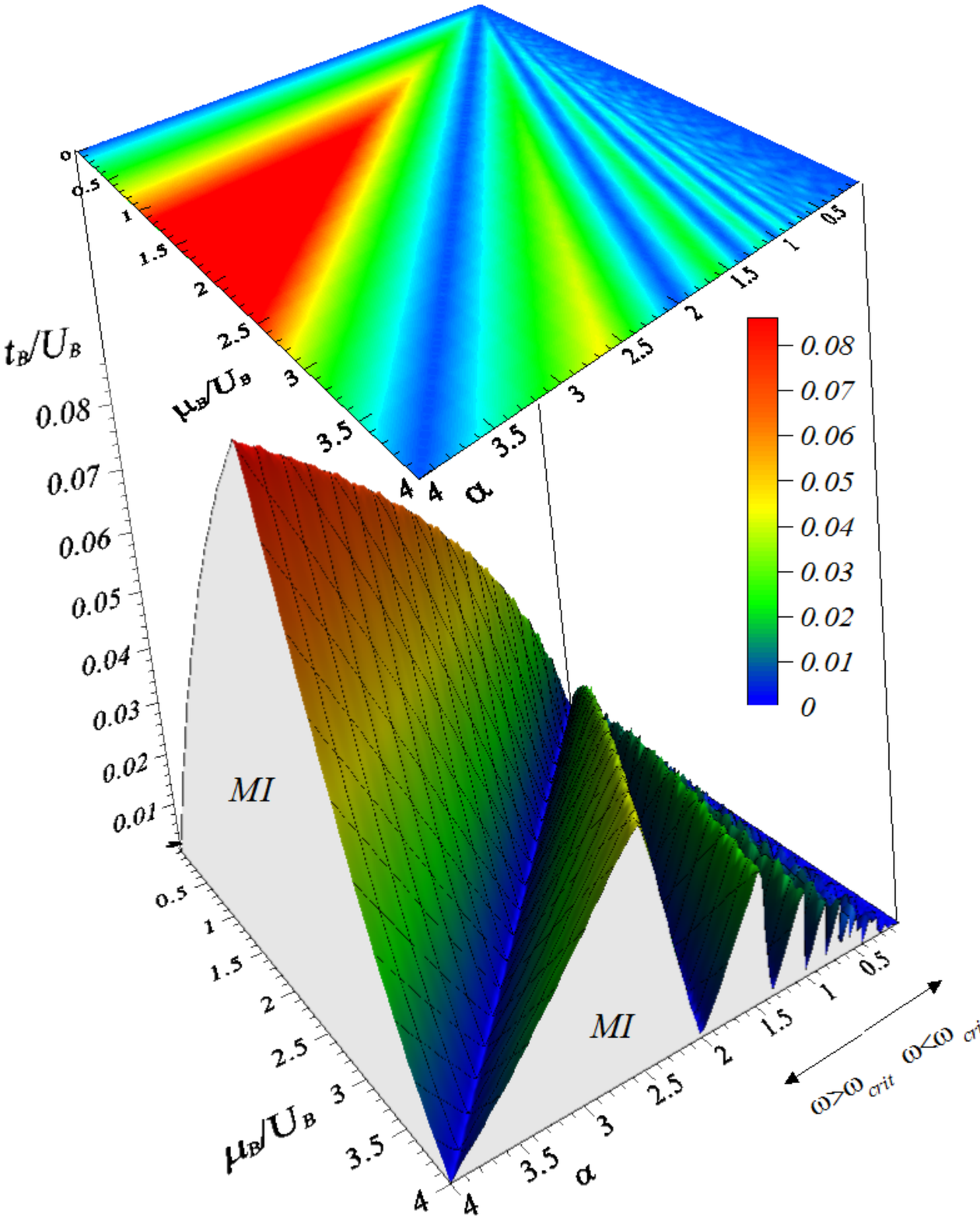}\caption{(Color online) The phase diagram showing the Mott insulator (MI) -
superfluid transition (MI phase inside the lobes, superfluid above
the surface) obtained from Eq. \ref{critical line final-1} of the
Bose-Fermi mixtures confined in two-dimensional square optical lattice
with synthetic magnetic field $f=1/2$ in the space of the parameters
$t_{b}/U_{b}$-$\mu_{b}/U_{b}$-$\alpha$ for $\eta=-2$ ($\omega>\omega_{\mathrm{crit}}$
means that system is in the repulsive regime $U_{\mathrm{eff}}>U_{b}$
with $\chi'\left(\omega\right)>0$, even though the interspecies interaction
is negative $U_{bf}<0$, see also Fig. \ref{Fig1}). Upper panel is
the density plot of the three-dimensional phase diagram.}
\label{Fig3}
\end{figure}

The critical line equation that separates the Mott insulator - superfluid
transition Fig. \ref{Fig3}, details of similar derivation of the
critical line equation are described in \cite{polak2}, will take
simple form: \begin{equation}
1=\int_{-\infty}^{+\infty}\frac{\rho^{p/q}\left(\xi\right)d\xi}{\sqrt{2\bar{\xi}\left(8\frac{t_{b}}{U_{b}}+\frac{\mu_{b}}{U_{b}}-\eta\right)\frac{1}{\alpha}\frac{t_{b}}{U_{b}}+\upsilon^{2}\left(\frac{1}{\alpha}\frac{\mu_{b}}{U_{b}}\right)}}\label{critical line final-1}\end{equation}
where $\bar{\xi}=\xi_{\mathrm{max}}^{p/q}-\xi$ and $\xi_{\mathrm{max}}^{p/q}$
is the maximum of the band spectrum. The renormalization parameters
are defined as: $\alpha=1+\frac{U_{bf}^{2}}{U_{b}}\chi'\left(\omega\right)$
and $\eta=\frac{U_{bf}}{U_{b}}N_{\mathrm{F}}-1/2$. In Eq. (\ref{critical line final-1})
$\upsilon\left(\mu/U\right)=\mathrm{frac}\left(\mu/U\right)-1/2,$
where $\mathrm{frac}\left(x\right)$ is the fractional part of the
number. Because the higher values of the normalized chemical potential
for the fermions $\mu_{f}/t_{f}$ decreases $\chi'\left(\omega\right)$
and $\chi''\left(\omega\right)$ \cite{polak2}, consequently terms
containing explicitly the average density of fermions $N_{\mathrm{F}}$
will acquire more significance than these with exclusively the inter-species
interaction $U_{bf}$. The periodicity of the phase diagram can be
easily characterized by its evolution with changing the $\alpha$
parameter and applied magnetic field (see Fig. \ref{Fig3}). For $f=0$,
taking $\omega<\omega_{\mathrm{crit}}$, we recover the previous theoretical
results in which, after adding fermions to the system, the effective
interaction $U_{\mathrm{eff}}\left(\omega\right)$ becomes smaller
than repulsive energy $U_{b}$ for bosons only and superfluid phase
increases. However, in the local dynamic limit, when $\omega>\omega_{\mathrm{crit}}$
the Mott insulator phase becomes stronger and bosons tend to localize
on the lattice sites. Imprinting a phase factor on neutral particles
$f\neq0$ can provide very complex behavior of the mixtures and induced
effective interaction $U_{\mathrm{eff}}$ \emph{can oscillate} from
attractive to repulsive depending on frequency. The measurements by
the Bragg spectroscopy method of the linear response of correlated
two-dimensional BF mixtures at non-zero momentum transfer can give
us more insight into the excitations spectra modified by the applied
synthetic magnetic field. %

In conclusion, we have studied a planar mixture of bosons and spinless
fermions with synthetic magnetic field imposed on the system. We found
that the underlying dynamics of mixture of particles with different
statistics and masses entering a quantum-Hall regime is very complex
allowing effective bosonic interaction to be switched between repulsive
and attractive. The experimental evidence of our findings is feasible
however precise measurements of the magnetic field are requisite which
is possible with the recently developed optically synthesized magnetic
field for neutral atoms \cite{lin}.

We thank T. K. Kope\'c, R. Micnas and I. Spielman for discussion
and comments regarding the paper.

\end{document}